\documentclass[aoas,nameyear,MSNbibl,dvips]{arximspdf}
\usepackage{mathbh}
\usepackage{graphicx}

\doi{10.1214/10-AOAS398M}
\referstodoi{10.1214/10-AOAS398}
\volume{5}
\issue{1}
\pubyear{2011}
\firstpage{47}
\lastpage{51}

\makeatletter
\newcommand{\tx}{\tilde{x}}
\newcommand{\tX}{\widetilde{X}}
\newcommand{\tXt}{\widetilde{X}^T}
\newcommand{\abs}[1]{{\vert #1 \vert}}
\newcommand{\E}{\mathbf{E}}
\newcommand{\txt}{\tilde{x}^T}
\makeatother

\begin{document}
\begin{frontmatter}

\title{Discussion of:  A statistical analysis of multiple temperature proxies: Are
reconstructions of surface temperatures over the last\\ 1000 years reliable?\thanksref{T1}}
\runtitle{Discussion}
\pdftitle{Discussion on A statistical analysis of multiple temperature proxies:
Are reconstructions of surface temperatures over the last 1000 years reliable?
by B. B. McShane and A. J. Wyner}
\thankstext{T1}{Lamont--Doherty Earth Observatory contribution number 7438.}

\begin{aug}
\author{\fnms{Alexey} \snm{Kaplan}\corref{}\thanksref{t1}\ead[label=e1]{alexeyk@ldeo.columbia.edu}%
\ead[label=u1,url]{http://rainbow.ldeo.columbia.edu/\textasciitilde alexeyk}}

\thankstext{t1}{Supported by grants from the NSF (ATM-0902436), NOAA (NA07OAR4310060), and NASA
(NNX09AF44G).}

\runauthor{A. Kaplan}

\affiliation{Lamont--Doherty Earth Observatory of Columbia University}

\address{Lamont--Doherty Earth Observatory\\
61 Route 9W \\
P.O. Box 1000\\
Palisades, New York 10964\\
USA\\
\printead{e1}\\
\printead{u1}} 
\end{aug}

\received{\smonth{10} \syear{2010}}
\revised{\smonth{11} \syear{2010}}



\end{frontmatter}

\setcounter{footnote}{2}

McShane and Wyner (\citeyear{MW2011}) (hereinafter MW2011) demonstrated that in many
cases a comprehensive data set of $p=1138$ proxies [Mann et al. (\citeyear{Metal2008})]
did not predict Northern Hemisphere (NH) mean temperatures
significantly better than random numbers. This fact is not very
surprising in itself: the unsupervised selection of good predictors
from a set of $p \gg  n$ proxies of varying sensitivities might be
too challenging a task for any statistical method ($p/n_c  \approx10$;
only $n_c=119$ out of total $n=149$ years were used for calibration in
MW2011 cross-validated reconstructions).  However, some types of
noise\footnote{Pseudoproxies used by MW2011 are called ``noise'' here;
  in climate research, pseudoproxies are synthetic combinations of a
  climate signal with some noise; without the former, it is a pure
  noise.} systematically outperformed the real proxies (see two bottom
panels of MW2011, Figure 10). This finding begs further investigation:
what do these random numbers have that real proxies do not?

To investigate this question, the present analysis uses ridge
regression [RR, Hoerl and Kennard (\citeyear{HK1970})] instead of the
Lasso [\citet{T1996}].\footnote{The difference is in the penalty norm: Lasso uses
  $L_1$ while RR uses $L_2$. MW2011 have also argued that a rough
  performance similarity should exist between different methods for
  $p \gg  n$ problems.}  The regression model used by MW2011 with Lasso and here
with RR is
\[
y=X\beta +
\beta_0\mathbh{1}_n+\varepsilon,
\]
where $y$ is a column vector of $n$ observations (annual NH
temperatures), $\varepsilon$ is random error, $X$ is a known $n \times
p$ matrix of predictors (climate proxies). A vector of regression
coefficients $\beta$ and an intercept constant $\beta_0$ are to be
determined. A~column $n$-vector $\mathbh{1}_n$ has all components
equal one. Proxy records are standardized before use; in
cross-validation experiments standardization is repeated for each
calibration period.

Let $w$ be a column $n_c$-vector such that
$w^T\mathbh{1}_{n_c}=1$. Define matrix-valued functions
$\mathcal{W}[w]=I-\mathbh{1}_{n_c}w^T$ and
$\mathcal{R}[S,\lambda,w]=S_{vc}(S_{cc}+\lambda
I)^{-1}\mathcal{W}[w]+\mathbh{1}_{n_v}w^T,$ where $S$ is a positive semidefinite
$n \times  n$ matrix, $\lambda > 0$ is the ridge parameter found
as a minimizer of the generalized cross-validation function [GCV,
Golub et al. (\citeyear{GHW1979})], matrix (or vector) subscripts $c$ or $v$
hereinafter indicate submatrices corresponding to the calibration or
validation periods, respectively. The RR reconstruction $\hat{y}_v$ of
temperatures in the validation period (a ``holdout block'' of $n_v=30$
consecutive years) is a linear transformation:
$\hat{y}_v=\mathcal{R}[S_p,\lambda,e]y_c$, where $S_p=\tX\tXt /  p$,
$\tX$ is the standardized version of $X$, and
$e={n}_c^{ -1}\mathbh{1}_{n_c}$.

\begin{figure}[b]
\vspace{-9pt}
\includegraphics{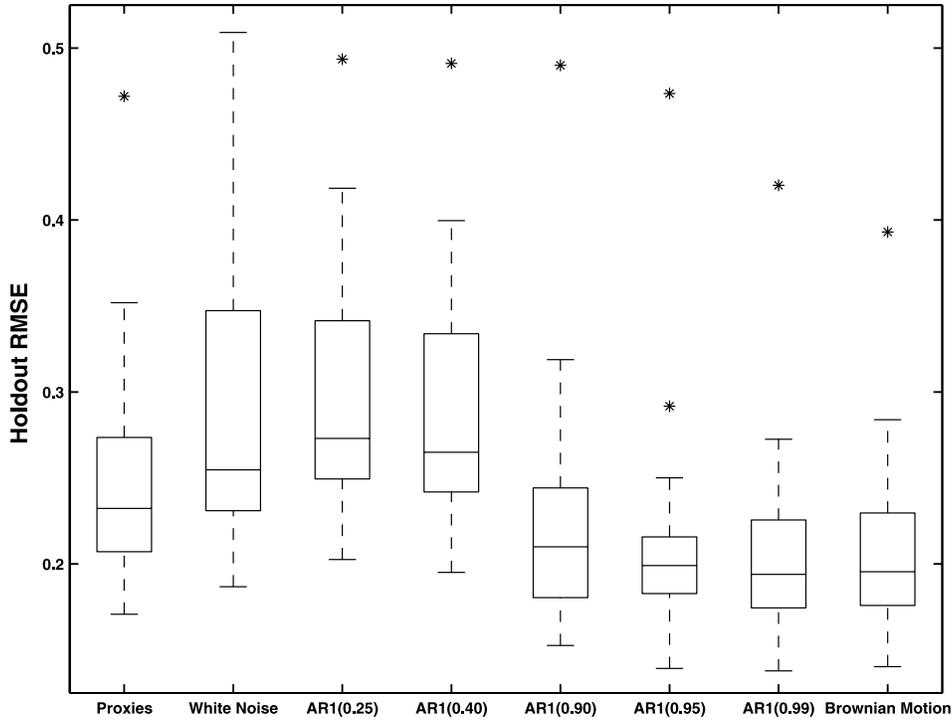}

\caption{Cross-validated RMSE on 120 30-year holdout blocks for the RR
  reconstructions from real climate proxies and from the random noise
  (one realization for each noise experiment); cf. MW2011, Figure~9.}\label{fig1}
\vspace{-15pt}
\end{figure}

\begin{figure}
\vspace{-9pt}
\includegraphics{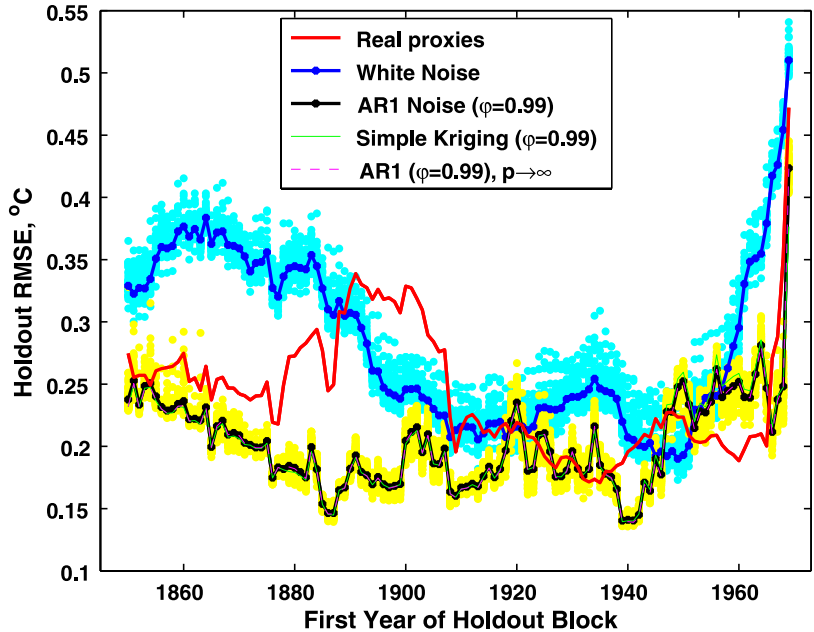}

\caption{Holdout RMSE for RR reconstructions as a function of time for
  real proxies (red) and two $100$-member ensemble means: white noise
  (blue) and $\operatorname{AR}(1)$ noise with $\varphi=0.99$ (black).
  The probability limit $(p\rightarrow\infty)$ for the
  latter is shown by magenta dashes. Holdout RMSE for simple kriging
  of the NH mean temperature index using an exponential semivariogram
  [Le and Zidek (\protect\citeyear{LZ2006})]
  $\gamma(\tau)=\lambda_{\min}+1-\exp[\tau\ln\varphi]$ with the
  GCV-selected nugget $\lambda_{\min}=\ell(\Phi,0)$ and long
  decorrelation scale $-1 / \ln(\varphi)=99.5$ years ($\tau$ is time
  in years) is shown by the green line. Individual
  ensemble members are shown by magenta and yellow dots,
  respectively.}\label{fig2}
  \vspace{-9pt}
\end{figure}

Using these formulas, the RR version of the MW2011 cross-validation
tests were performed for real proxies and for some noise
types. Results are shown in Figure~\ref{fig1}. The cross-validated root mean square error (RMSE) of the
RR reconstructions are smaller than Lasso values (cf. MW2011, Figure~9),
but the relative performance in different experiments appears
consistent between RR and Lasso. As in the Lasso case, noise with high temporal persistence, that is, simulated by the Brownian
motion or by the first-order autoregressive process $\operatorname{AR}(1)$ with a parameter $\varphi\geq0.9$,
outperformed proxies. Figure \ref{fig2} illustrates the time dependence of
the holdout error for the real-proxy, white-noise, and $\varphi=0.99$
AR(1) cases.  There is a general similarity between these and the
corresponding curves in Figure 10 by MW2011.

Note that a traditional approach to
hypothesis testing would evaluate an RMSE corresponding to a
regression of temperature data ($y$) on real proxies ($X$) in the
context of the RMSE probability distribution induced by the assumed
distribution of $y$ under the hypothesized condition (e.g.,
$\beta=0$). However, MW2011 evaluate the RMSE of real proxies in the
context of the RMSE distribution induced by random values in $X$, not
$y$. Such an approach to testing a null hypothesis would be
appropriate for an inverse relationship, that is,
$X=y\beta^T+\mathbh{1}_n\beta_0^T+\varepsilon$. When used with a
direct regression model here, however, it results in the RMSE
distribution with a surprising feature: when
$p\rightarrow\infty$, RMSE values for individual realizations of the
noise matrix $X$ converge in probability to a constant.

This convergence occurs because the columns $x$ of $X$ in the noise
experiments are i.i.d. from the noise distribution; AR(1) with
$\varphi=0.99$ is considered here: $x\sim\mathcal{N}(0,\Phi),
\Phi=(\varphi^\abs{i-j})$. The columns of $\tX$ are i.i.d. too, hence
the random matrix $S_p=\tX\tXt /p$ is an average of $p$
i.i.d. variates $\tx\tx^T$. Expectation $\Psi=\E \tx\txt$ exists; its
elements are computed as expectations of ratios and first inverse
moments of quadratic forms in normal variables [Jones (\citeyear{J1986}, \citeyear{J1987})].\vspace*{-2pt}
The weak law of large numbers applies, so
$S_p\stackrel{P}{\rightarrow}\Psi$. Since the GCV function depends on
$S$ and $w$ as well as on $\lambda$, its minimizing $\lambda$ will
depend on these parameters too: $\lambda_{\min}=\ell[S,w]$. Here GCV
is assumed well-behaved, so that $\ell$ is a single-valued function,
continuous at $(\Psi,e)$. From the definition of $\mathcal{R}$,
$\mathcal{B}[S,e]\equiv\mathcal{R}[S,\ell[S,e],e]$ will also be continuous at
$S=\Psi$, thus $S_p\stackrel{P}{\rightarrow}\Psi$ implies
$\hat{y}_v=\mathcal{B}[S_p,e]y_c\stackrel{P}{\rightarrow}\mathcal{B}[\Psi,e]y_c$.

When $p$ is finite but large, like $p=1138$, reconstructions based on
individual realizations of a noise matrix $X$ are dominated by their
constant components, especially when $\varphi\approx 1$: note the
small scatter of RMSE values in the ensemble of AR(1) with
$\varphi=0.99$ (yellow dots in Figure \ref{fig2}). The probability limit
$\hat{y}_v=\mathcal{B}[\Psi,e]y_c$ yields RMSE values (magenta dash in
Figure \ref{fig2}) that are very close ($1.3{\cdot} 10^{-3 \,\circ}$C RMS
difference) to the ensemble mean RMSE (black curve in Figure \ref{fig2}). To
interpret this non-random reconstruction, consider its simpler
analogue, using neither proxy standardization nor a regression
intercept ($\beta_0$).  Then, if the assumptions on the GCV function
change accordingly,
$\hat{y}_v\stackrel{P}{\rightarrow}\mathcal{B}[\Phi,0]y_c=\Phi_{vc}[\Phi_{cc}+\ell(\Phi,0)I]^{-1}y_c$,
that is, a prediction of $y_v$ from $y_c$ by ``simple kriging'' [Stein
(\citeyear{S1999}, page 8)], which in atmospheric sciences is called objective analysis
or optimal interpolation [Gandin (\citeyear{G1963})]. The RMSE corresponding to
this solution is shown in Figure \ref{fig2} (green line): it is quite close to the ensemble
mean RMSE for AR(1) noise with $\varphi=0.99$ (RMS difference is
$5.4{\cdot} 10^{-3 \,\circ}$C). The solution $\mathcal{B}[\Psi,e]y_c$, to
which the noise reconstructions without simplifications converge as
$p\rightarrow\infty$, is more difficult to interpret. Still, it has a
structure of an objective analysis solution and gives results that are similar
 to simple kriging: the RMS difference between the two
reconstructions over all holdout blocks is $7.7{\cdot} 10^{-3 \,\circ}$C.

Due to the large value of $p$ in the MW2011 experiments, their tests
with the noise in place of proxies essentially reconstruct holdout
temperatures by a kriging-like procedure in the temporal
dimension. The covariance for this reconstruction procedure is set by
the temporal autocovariance of the noise.  Long decorrelation scales
($\varphi\ge 0.95$) gave very good results, implying that long-range
correlation structures carry useful information about predictand time
series that is not supplied by proxies.  By using such a noise for
their null hypothesis, MW2011 make one skillful model (multivariate
linear regression on proxies) compete against another (statistical
interpolation in time) and conclude that a loser is useless. Such an
inference does not seem justified.

Modern analysis systems do not throw away observations simply because
they are less skillful than other information sources: instead, they
combine information.  MW2011 experiments have shown that their
multivariate regressions on the proxy data would benefit from
additional constraints on the temporal variability of the target time
series, for example, with an AR model. After proxies are combined with such a
model, a test for a significance of their contributions to the common
product could be performed.

\section*{Acknowledgements}
Generous technical help and many useful comments from Jason Smerdon
and very helpful presentation style guidance from Editor Michael Stein
are gratefully acknowledged.

%

\begin{supplement}[id=suppA]
\stitle{Data and codes}
\slink[doi]{10.1214/10-AOAS398MSUPP}  
\slink[url]{http://lib.stat.cmu.edu/aoas/398M/supplementM.zip}
\sdatatype{.zip}
\sdescription{This supplement contains a tar archive with all data files and codes (Matlab scripts) needed for
reproducing results presented in this discussion. Dependencies between files in the archive and
the order in which Matlab scripts have to be executed are described in the file \textit{README\_final},
also included into the archive.}
\end{supplement}


\printaddresses

\end{document}